\documentclass[12pt,preprint]{aastex}

\newcommand{\dpar}[2][\;\;]{\ensuremath{ \frac{\partial{#1}}{\partial{#2}} }}
\newcommand{\dparn}[3][\;\;]{\ensuremath{ \frac{\partial^{#3}{#1}}{\partial{#2}^{#3}} }}

\newcommand{\va}{V_{\rm A}}

\newcommand{\bvec}[1]{{\mbox{{\boldmath$#1$}}}} 
\newcommand{\unitv}[1]{\bvec{\hat{#1}}}

\newcommand{\eqnref}[1]{(\ref{#1})}

\begin{document}

\title{An Interpretation of Flare-Induced and Decayless Coronal-Loop Oscillations as Interference Patterns}

\author{Bradley W. Hindman}
\affil{JILA and Department of Astrophysical and Planetary Sciences,
University of Colorado, Boulder, CO~80309-0440, USA}

\author{Rekha Jain}
\affil{School of Mathematics \& Statistics, University of Sheffield, Sheffield S3 7RH, UK}

\email{hindman@solarz.colorado.edu}


\begin{abstract}

We present an alternative model of coronal-loop oscillations which considers that
the waves are trapped in a 2D waveguide formed by the entire arcade of field lines.
This differs from the standard 1D model which treats the waves as the resonant
oscillations of just the visible bundle of field lines. Within the framework of
our 2D model, the two types of oscillations that have been observationally identified,
flare-induced waves and ``decayless" oscillations, can both be attributed to MHD
fast waves. The two components of the signal differ only because of the duration
and spatial extent of the source that creates them. The flare-induced waves are
generated by strong localized sources of short duration, while the decayless
background can be excited by a continuous, stochastic source. Further, the oscillatory
signal arising from a localized, short-duration source can be interpreted as a
pattern of interference fringes produced by waves that have traveled diverse routes
of various pathlengths through the waveguide. The resulting amplitude of the fringes
slowly decays in time with an inverse square root dependence. The details of the
interference pattern depend on the shape of the arcade and the spatial variation
of the Alfv\'en speed. The rapid decay of this wave component, which has previously
been attributed to physical damping mechanisms that remove energy from resonant
oscillations, occurs as a natural consequence of the interference process without
the need for local dissipation.

\end{abstract}



\section{Introduction}
\label{sec:introduction}
\setcounter{equation}{0}

The detection of standing kink-wave oscillations on bright coronal loops by the
TRACE instrument \citep[e.g.,][]{Aschwanden:1999, Nakariakov:1999} proffered the
possibility that seismic techniques could be applied to magnetic structures in
the corona. Given that the magnetic field of the corona remains resistant to
measurement by spectroscopic means, coronal-loop seismology permits the direct
probing of an otherwise inaccessibile (yet paramount) property of the corona.
Two observational details suggest that many of the observed motions are those
arising from resonant standing waves with wavelengths corresponding to the
fundamental mode. These observational details are that different segments of the
loop are often observed to vascillate in phase with each other and the sinuous
motion usually lacks nodes, except perhaps at the loop footpoints. In some instances
in addition to the fundamental mode with the lowest frequency, coexistent overtones
(with interior nodes and a higher frequency) have been detected \citep{Verwichte:2004,
VanDoorsselaere:2007, DeMoortel:2007}. The discovery of such overtones precipitated
a tumult of theoretical activity with the goals of both explaining the observed
dispersion and developing seismic methods that use the dispersion to surmise the
field strength and mass density along the loop \citep[see the review by][]{Andries:2009}.

The first oscillations to be detected were the response of loops within coronal
arcades to the passage of transient disturbances launched from solar flares. It
has been observed that such oscillations, once initiated, rapidly diminish over
3--4 wave periods \citep[e.g.,][]{White:2012}. A variety of theoretical studies
have suggested possible damping mechanisms to explain the observed diminuation
of the signal, with the most prominent being resonant absorption
\citep[e.g.,][]{Ruderman:2002, Goossens:2002, Goossens:2011} and phase mixing
between distinct fibrils in a bundle, each with slightly different wave speeds
\citep{Ofman:2002}. In all cases the diminuation of signal results from a physical
loss of energy from the observed kink waves.

Recent observations \citep{Nistico:2013, Anfinogentov:2013} using the Atmospheric
Imaging Assembly (AIA) on the Solar Dynamics Observatory (SDO) have revealed that
in addition to these large-amplitude flare-induced oscillations there appears to
be a continuous background of fluctuating power that oscillates at frequencies
similar to the flare-induced waves, but with a lower amplitude that does not exhibit
significant attenuation. These studies have posited that the background oscillations
are excited by a continuous, and perhaps stochastic, driver whose energy input is
balanced on the long term by physical damping. Thus, the two classes of oscillation
are caused by waves with the same resonant nature, but excited by different sources,
one ongoing and the other impulsive.

Implicit in much of this prior work is the assumption that each visible loop in 
a coronal arcade is essentially an independent wave cavity. The MHD kink waves
are presumed to have a group velocity that is parallel to the field lines and
each loop oscillates as a coherent entity. Thus, the problem is reduced to a
1D wave problem and any decay in the signal must be due to a physical loss of
energy.  Here we explore an alternative interpretation. Two properties of the
observed oscillations suggest that the wave cavity may be inherently 2D instead
of 1D. The first of these properties is that most oscillating loops appear to be
part of a coronal arcade and that flares often cause the entire arcade to ring
and throb. The second property is that most of the standing waves that have been
observed possess a ``horizontal" polarization such that the loop sways back and
forth within the arcade without significant expansion and contraction of the loop's
radius. This implies that the motion of a loop impacts neighboring (and possibly
invisible) field lines within the arcade and forces them to move.  The individual
bundles of field lines that form loops are therefore not isolated from the larger
arcade structure in which they are embedded. Previously, the ringing of the arcade
has been interpreted to be the transient disturbance, launched by the flare, that
excites the resonant oscillations separately on each field line. Here we suggest
that the arcade forms a 2D waveguide and that the ringing is a superposition of
waveguide modes that form in response to driving by flares and other sources.
The waves are inherently 2D modes that are trapped standing waves longitudinal
to the field, while propagating up and down the axis of the arcade perpendicular
to the field. We will find that such a model naturally leads to decaying signals
without the need for a physical damping mechanism. The attenuation of the sinusoidal
signal observed at a given field line or loop is a fringe pattern resulting from
the self interference of a wavefront as it expands away from a point source.

In Section~\ref{sec:Waveguide} we derive a simple wave equation that describes
MHD fast waves that propagate on a thin 2D sheet of arching field lines. We define
the geometry of the field and describe the boundary conditions that turn the arcade
into a waveguide. In Section~\ref{sec:Signal} we derive the response of the waveguide
to two types of sources, a continuous stochastic source and an impulsive source.
Finally, in Section~\ref{sec:Discussion} we discuss the implications of our calculation
and present our conclusions.


\section{MHD Fast Waves in a Waveguide}
\label{sec:Waveguide}

We treat a coronal arcade as a thin magnetized sheet with each field line in the
sheet piercing the photosphere at two locations. The locus of the footpoints for
all of the field lines form two parallel lines in the photosphere. If we view the
photosphere as a dense, immovable fluid, waves are trapped between the footpoints
and the arcade acts as a 2D waveguide for MHD fast waves that permits free propagation
up and down its axis perpendicular to the field. We utilize a Cartesian coordinate
system oriented such that the $x$--$y$ plane corresponds to the photosphere and
the $z$ coordinate is the height above the photosphere. Let the arcade be invariant
in the $y$-direction and let it lack sheer such that the field has no component in
that direction,

\begin{equation}
	\bvec{B} = B_x(x,z) \unitv{x} + B_z(x,z) \unitv{z} \; .
\end{equation}

\begin{figure*}%
        \epsscale{1.0}%
        \plotone{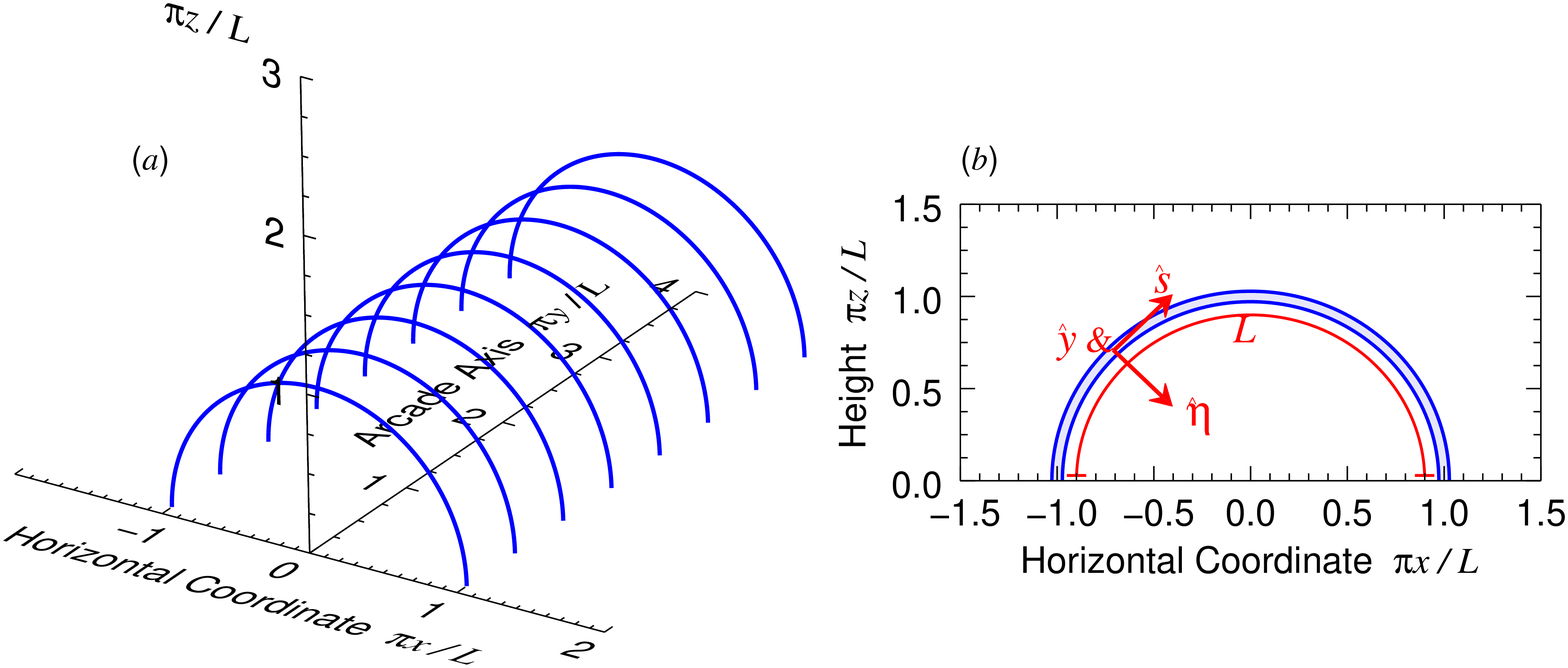}%
        \caption{\small Schematic diagram of a coronal arcade. The $x$-$y$ plane
corresponds to the photosphere and the height above the photosphere is given by
$z$. ($a$) Three-dimensional view of the thin sheet of magnetic field lines that
define the arcade. The arcade lacks shear and is invariant along the axis of the
arcade in the $y$-direction. ($b$) Cross-sectional cut through the arcade at constant
$y$ showing a thin annulus. The triad of unit vectors for the local Frenet coordinates
are shown in red, with the tangent vector $\unitv{s}$, the principal normal $\unitv{\eta}$,
and the binormal $\unitv{y}$. Each field line has a length of $L$ from photosphere
to photosphere. For the sake of presentation, we assume that each field line forms
a semicircle.
\label{fig:Schematic}}%

\end{figure*}

\noindent Figure~\ref{fig:Schematic} provides a schematic diagram of the arcade
and its geometry.
 
For simplicity, we will assume that the corona is magnetically dominated such
that gravity and gas pressure can be ignored when compared to the magnetic forces.
Furthermore, in order to build an illustrative example without unnecessary mathematical
complication, we will ignore the curvature of the field lines within the wave
equation and assume that the Alfv\'en speed $\va$ within the sheet is uniform.
With these assumptions, fast MHD waves can be conveniently expressed using local
Frenet coordinates. The triad of unit vectors that represents this local coordinate
system are the tangent to the field line $\unitv{s}$, the field line's principal
normal $\unitv{\eta}$, and the binormal $\unitv{y}$. The tangential coordinate
$s$ measures the pathlength along a field line starting from the photosphere,
with the footpoints located at $s=0$ and $s=L$, where the length of each field
line $L$ is the same. The coordinate $y$ marks distance along the axis of the
waveguide and we assume that the arcade is long enough that we can ignore edge
effects.


\subsection{Equation for Driven Fast Waves}
\label{subsec:FastWaves}

Since we are only considering magnetic forces, fast MHD waves lack motion parallel
to the field lines and the transverse motion is irrotational,

\begin{eqnarray}
	\bvec{v} &=& v \unitv{y} + w \unitv{\eta} \; ,
\\ \nonumber \\
	\dpar[v]{\eta} &=& \dpar[w]{y} \; .
\end{eqnarray}

\noindent Further, driven fast waves satisfy the following simple equation,

\begin{equation} \label{eqn:WaveEqn}
	\left[ \dparn[ ]{t}{2} -
		\va^2 \left(\dparn[ ]{y}{2} + \dparn[ ]{s}{2} + \dparn[ ]{\eta}{2}\right) \right] \bvec{v}
			= \bvec{S}(\bvec{x},t) \; ,
\end{equation}

\noindent where $\bvec{S}(\bvec{x},t)$ is a wave driver, the Alfven speed $\va$
is given by $\va^2 = B^2/4\pi\rho$, and $\rho$ is the mass density. In observational
contexts, motions in the direction of the principal normal, $w$,  are often referred
to as ``vertical" oscillations, whereas velocities $v$ in the binormal direction,
along the axis of the arcade, are called ``horizontal" oscillations. We will explore
horizontal oscillations ($w=0$) by supposing that the driver only acts in the
binormal direction and lacks variation across the arcade's sheet, i.e., the driver
is independent of the coordinate $\eta$. With these assumptions, only 2D wave modes
are excited and they obey the following equation:

\begin{equation} \label{eqn:WaveEqn_2D}
	\left[ \dparn[ ]{t}{2} -
		\va^2 \left(\dparn[ ]{y}{2} + \dparn[ ]{s}{2}\right) \right] v
			= S(s,y,t) \; .
\end{equation}

We make the arcade into a waveguide by imposing boundary conditions at the footpoints.
Specifically, we apply the line tieing condition at the photosphere (i.e., $v=0$
at $s=0$ and $s=L$). The resonant modes of this waveguide form a discrete spectrum
in the tangential $s$-direction and a continuous spectrum in the transverse $y$-direction
down the axis of the waveguide,

\begin{eqnarray}
	v_n(s,y,t; \kappa) &=& U_n(s) \, e^{i\kappa y} \, e^{-i \omega_{n}(\kappa) t} \; ,
\\ \nonumber \\
	U_n(s) &\equiv& \left(\frac{2}{L}\right)^{1/2} \, \sin\left(\lambda_n s\right) \; ,
\end{eqnarray}

\noindent for positive mode orders $n = 1, 2, 3, 4, \cdots$~. The allowed
parallel wavenumbers $\lambda_n$ and the eigenfrequency $\omega_n(\kappa)$ are
given by

\begin{eqnarray}
	\lambda_n &=& \frac{n \pi}{L} \; ,
\\ \nonumber \\
	\omega_{n}^2(\kappa) &=& \left(\lambda_n^2 + \kappa^2\right) \va^2 \; .
\end{eqnarray}

\begin{figure*}%
        \epsscale{0.5}%
        \plotone{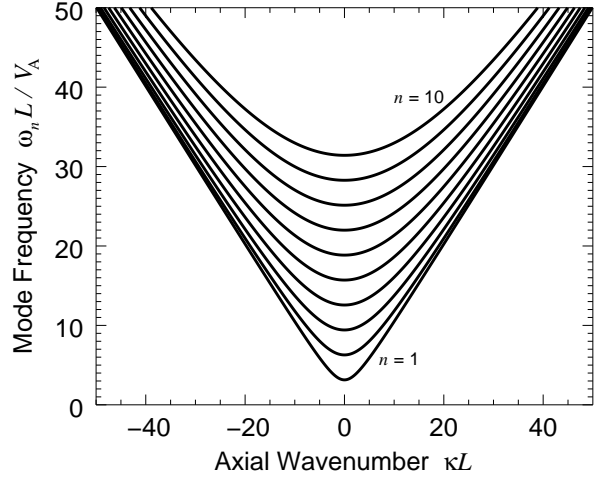}%
        \caption{\small The eigenfrequencies, $\omega_n^2 = \lambda_n^2 + \kappa^2$,
of the modes of the waveguide as a function of the wavenumber $\kappa$ parallel to
the waveguide's axis. Each curve corresponds to a different mode order $n$ labelling
the discretely allowed wavenumbers $\lambda_n = n \pi/L$ in the direction parallel
to the field. The gravest mode ($n=1$) has the lowest frequency, and each higher order
has a correspondingly higher frequency.
\label{fig:EigFreqs}}%

\end{figure*}

\noindent The wave is a standing wave in the direction parallel to the magnetic field,
with discrete wavenumbers $\lambda_n$, and a propagating wave in the $y$ direction with
continuous wavenumber $\kappa$. The temporal frequency $\omega_{n}(\kappa)$ depends
on both wavenumbers and is illustrated in Figure~\ref{fig:EigFreqs}. The parallel
eigenfunctions, $U_n(s)$, have been normalized such that they form an orthonormal set.

Our general strategy for solving the driven equation~\eqnref{eqn:WaveEqn_2D} is as
follows: we will Fourier transform the equation in the invariant $y$-direction,
decompose the source and solution into the eigenfunctions of the waveguide, solve
for the amplitude of each mode in spectral space, and then return to configuration
space by inverting the transform. After Fourier transforming the driven wave equation
and projecting onto the eigenmodes of the waveguide, we obtain

\begin{equation} \label{eqn:WaveEqnComp}
	\left[\dparn[ ]{t}{2} + \omega_{n}^2(\kappa)\right] \hat{v}_{n}(\kappa,t) = \hat{S}_{n}(\kappa,t) \; ,
\end{equation}

\noindent where

\begin{eqnarray}
	\hat{S}_{n}(\kappa,t) &=& \int_{-\infty}^\infty dy \int_0^L ds \; S(s,y,t) \, U_{n}(s) \, e^{-i\kappa y} \; ,
\\ \nonumber \\
	\hat{v}_{n}(\kappa,t) &=&  \int_{-\infty}^\infty dy \int_0^L ds \;
		v(s,y,t) \, U_{n}(s) \, e^{-i\kappa y} \; .
\end{eqnarray}

In configuration space, the solution is obtained by inverting the transform and summing
over eigenmodes,

\begin{equation}
	v(s,y,t) = \frac{1}{2\pi} \int_{-\infty}^\infty dy \sum_{n=1}^\infty \,
		\hat{v}_n(\kappa,t) \, U_{n}(s) \, e^{i\kappa y} \; .
\end{equation}

\noindent A similar equation holds for the reconstruction of the source from its
spectral decomposition,

\begin{equation}
	S(s,y,t) = \frac{1}{2\pi} \int_{-\infty}^\infty dy \sum_{n=1}^\infty \,
		\hat{S}_n(\kappa,t) \, U_{n}(s) \, e^{i\kappa y} \; .
\end{equation}


\section{Two-Component Signal}
\label{sec:Signal}

We posit that the wave signal seen at the observation location $y$ is a superposition
of the waves generated by a source with two components: a broad-band driver that
generates a low-amplitude, resonant, background signal and an energetic impulsive
source that generates a large initial pulse with subsequent ringing,

\begin{equation} \label{eqn:TwoSource}
	S(s,y,t) = S_{\rm bg}(s,y,t) + S_{\rm imp}(s) \delta(t-t^\prime) \delta(y-y^\prime) \; .
\end{equation}

\noindent The first term $S_{\rm bg}$ represents the continuous, broad-band driver,
which could be the incessant buffeting from ambient waves in the corona external to
the waveguide, or perhaps the random movement of the footpoints of the arcade in the
photosphere by convective motions. The second term $S_{\rm imp}$ is the impulsive
source arising from a single short duration event such as a flare. Of course each
source will independently produce a wave response,

\begin{equation}
	v(s,y,t) = v_{\rm bg}(s,y,t) + v_{\rm imp}(s,y,t) \; .
\end{equation}


\subsection{Resonant Background Oscillations}
\label{subsec:Background}

The background velocity resulting from the broad-band component of the source can
be expressed as a superposition of waveguide modes. The amplitude and phase of each
mode can be obtained by taking the temporal Fourier transform of equation~\eqnref{eqn:WaveEqnComp},
solving for the velocity in spectral space, and inverting the temporal transform
through contour integration (see Appendix A),

\begin{equation} \label{eqn:v_bg}
	v_{\rm bg}(s,y,t) = \frac{1}{2\pi} \sum_{n=1}^\infty \int_{-\infty}^\infty d\kappa \;
		A_n(\kappa) \,U_n(s) \, \sin\left[\kappa y - \omega_n(\kappa) t + \theta_n(\kappa)\right] \; .
\end{equation}

\noindent In this equation, $A_n(\kappa)$ is the mode amplitude of the mode and
$\theta_n(\kappa)$ is the phase of the source function in spectral space, evaluated
at the mode frequencies,

\begin{eqnarray}
	\hat{S}^{\rm (bg)}_{n}(\kappa,\omega) &=& \int_{-\infty}^\infty dt \int_{-\infty}^\infty dy \int_0^L ds \; S_{\rm bg}(s,y,t) \, U_{n}(s) \, e^{-i(\kappa y-\omega t)} \; ,
\\ \nonumber \\
	\hat{S}^{\rm (bg)}_{n}(\kappa,\omega_n) &=& \left|\hat{S}^{\rm (bg)}_{n}(\kappa,\omega_n)\right| e^{i\theta_n(\kappa)} \; ,
\\ \nonumber \\
	A_n(\kappa) &\equiv& \frac{\left|\hat{S}^{\rm (bg)}_{n}(\kappa,\omega_n)\right|}{\omega_n(\kappa)} \; .
\end{eqnarray}

\noindent The amplitude of the mode $A_n(\kappa)$ depends on the modulus of the source
function evaluated at the mode frequency. The factor of frequency appearing in the
denominator of the mode amplitude is a direct result of the fact that a white source
excites modes such that they all have equal energy. Since the energy in each mode
$E_n$ is proportional to both the square of the amplitude $A_n(\kappa)$ and the square
of the frequency $\omega_n(\kappa)$, we should expect the amplitude of each mode to
be inversely proportional to its frequency. Even if the source function possesses
wavenumber dependence, we still expect the waves with the lowest frequency to dominate
the background signal as long as the source is not a rapidly increasing function of
wavenumber.

The phase of the response depends on the phase $\theta_n$ of the source function and
the variation of this phase over all wavenumbers $\kappa$ comprising the signal. Since
the signal has contribution from a range of frequencies around a dominant frequency,
we expect the interference between the different frequencies to cause beat patterns
that will slowly and randomly rotate the apparent phase of the oscillation in time.
Therefore, at a given position along the waveguide, we should expect the resonant
background to be dominated by the gravest mode and produce a signal with the following
form,

\begin{equation}
	v_{\rm bg} \approx A_{\rm bg} \, \sin(\pi s/L) \, \sin\left[(\pi \va/L) t + \phi(t)\right] \; ,
\end{equation}

\noindent where the phase $\phi(t)$ slowly changes with time ($|\dot{\phi}_n| << \pi \va/L$)
due to the continual excitation by the source.

\begin{figure*}%
        \epsscale{1.0}%
        \plotone{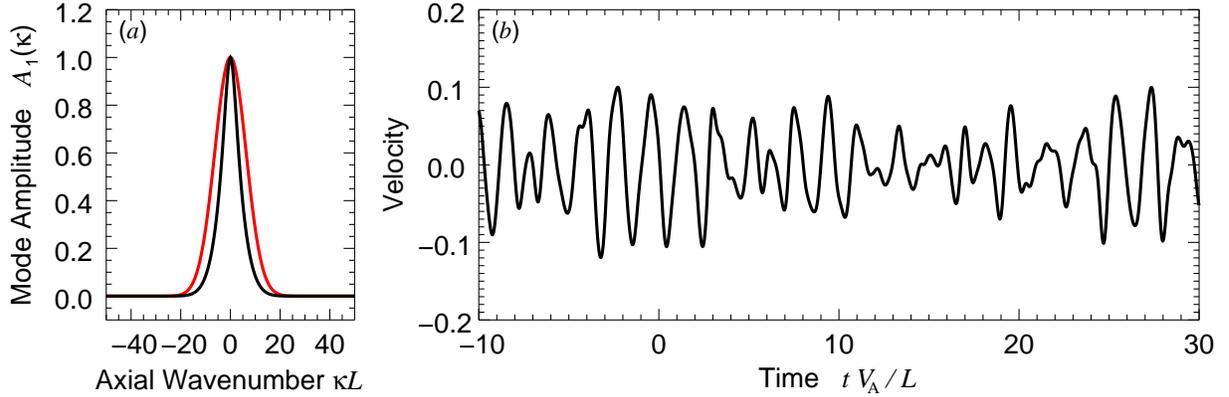}%
        \caption{\small The background oscillations described both in spectral space
and as a function of time. The source strength $|\hat{S}^{\rm (bg)}_{1}|$ is a Gaussian
function of wavenumber, chosen to have unit amplitude ($\tilde{S}=1$) and a width of
$\Delta = 2\pi/L$. ($a$) The source strength (red curve) and the resulting amplitude
spectrum $A_1(\kappa) = |\hat{S}^{\rm (bg)}_{1}|/\omega_1(\kappa)$ (black curve) for
the gravest mode ($n=1$) as a function of axial wavenumber $\kappa$. ($b$) The time
series of the background oscillation as observed at the apex of the arcade, $s = L/2$,
and at an arbitrary position along the arcade $y=y_{\rm obs}$. The signal has a dominant
frequency $\omega = n\pi\va/L$, with a phase that slowly wanders with time.
\label{fig:GaussSpec}}%

\end{figure*}

We will see that this expectation holds true by exploring in more detail two types
of background sources. In Figure~\ref{fig:GaussSpec}$a$ we show the source strength
(red curve) for a source with a Gaussian dependence on wavenumber,

\begin{equation}
	\left|\hat{S}^{\rm (bg)}_{n}(\kappa,\omega_n)\right| = \tilde{S} \, \exp\left(-\frac{\kappa^2}{2\Delta^2}\right) \delta_{n1} \; .
\end{equation}

\noindent In this equation, $\tilde{S}$ is an arbitrary constant and $\Delta$ is the
spectral width of the Gaussian. For simplicity we have assumed that the source only
excites the fundamental mode $n=1$. We model a stochastic source by imposing that the
phase of the source $\theta_n(\kappa)$ is a random function with a uniform distribution
between $0$ and $2\pi$. Figure~\ref{fig:GaussSpec}$a$ also shows the mode amplitude
$A_n(\kappa)$ as the black curve. This type of source generates an amplitude spectrum
that is sharply peaked at zero wavenumber with little contribution from the wings.
Thus we expect that the corresponding time-series (shown in Figure~\ref{fig:GaussSpec}$b$)
should be dominated by the frequency $\omega_1 = \pi \va/L$, with a phase that slowly
wanders. This is indeed the case. This time-series was constructed by numerically
evaluating the inverse spatial transform in Equation~\eqnref{eqn:v_bg}.

\begin{figure*}%
        \epsscale{1.0}%
        \plotone{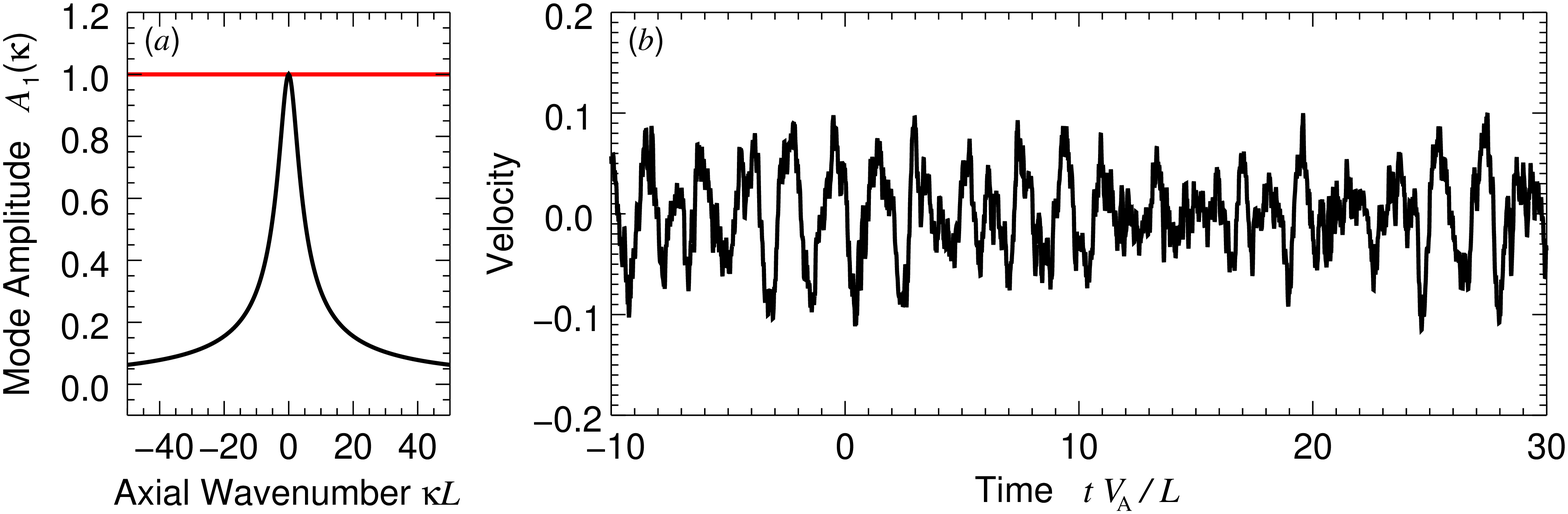}%
        \caption{\small The background oscillations described both in spectral space
and as a function of time for a source that is white with a strength that is independent
of wavenumber. ($a$) The mode amplitude $A_1(\kappa)$ (black curve) of the background
oscillations for the gravest mode ($n=1$) as a function of axial wavenumber $\kappa$.
The source strength is overlayed in red. Even though the source is white, the signal
is dominated by the waves with the smallest axial wavenumbers and hence the waves with
the lowest frequency (see Figure~\ref{fig:EigFreqs}). However, the wings of the amplitude
distribution are more significant than they are for the Gaussian source. ($b$) The
time series of the background oscillation as observed at the apex of the arcade, $s = L/2$,
and at an arbitrary point along the arcade $y = y_{\rm obs}$. Since the wings are enhanced
in the amplitude distribution, the time series has more prominant high-frequency jitter.
\label{fig:WhiteSpec}}%

\end{figure*}

\begin{figure*}%
        \epsscale{0.5}%
        \plotone{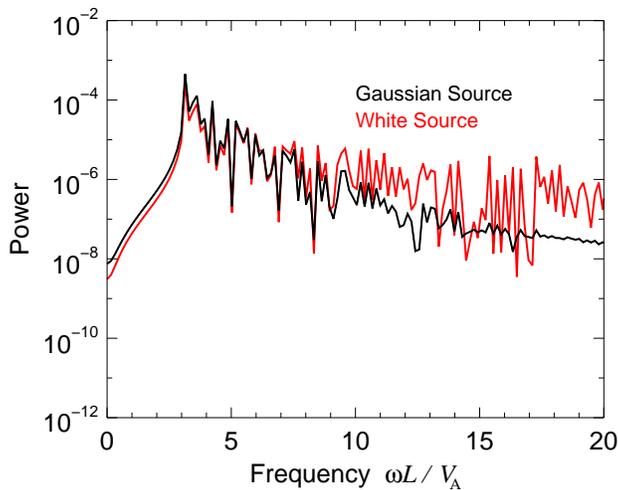}%
        \caption{\small Temporal power spectra of the background oscillations illustrated
in Figures~\ref{fig:GaussSpec} and \ref{fig:WhiteSpec}. The fine structure arises
from the interference between waves and is, therefore, sensitive to the specific
realization of wave phases. Here, for the sake of comparison, we have used the same
realization for both types of source. The spectrum generated by the white source
clearly has a greater contribution from high-frequency waves. Both spectra have a
low-frequency cut-off that corresponds to the resonant mode frequency $\omega_1 = \pi \va/L$
appropriate for propagation parallel to the field lines (i.e., $\kappa = 0$). 
\label{fig:PowSpec}}%

\end{figure*}

For the second source we choose a white source whose strength (by definition) is a
constant function of wavenumber. The phase of the source function is once again chosen
to be random. Figure~\ref{fig:WhiteSpec} presents the amplitude spectrum and the
resulting time series. Since the white source contains a broader range of wavenumbers,
it generates a time-series with richer frequency response. In particular, we can clearly
see from Figure~\ref{fig:WhiteSpec}$b$ that the time series possesses high-frequency
jitter. The time-series generated by both sources are highly correlated with very similar
low-frequency behavior. This is because the same realization of random phases was used
to construct both sources. The equivalency of the set of phases also manifests in the
temporal power spectra of the two time series (see Figure~\ref{fig:PowSpec}). The fine
structure in the two power spectra is similar because this structure arises from wave
interference, which of course is determined by the relative phases (which were chosen
to be identical).


\subsection{Response to an Impulsive Source}
\label{subsec:Impulsive}

The waves generated by the impulsive source are of course determined by the Green's
function. Therefore, consider a single point source of unit amplitude that occurs at
time $t^\prime$ and at location $(s,y) = (s^\prime, y^\prime)$. We perform a detailed
derivation of the Green's function in Appendix B. The general procedure is to Fourier
transform  the wave equation in the axial direction $y$, decompose into waveguide
modes, and solve for the temporal behavior. The solution is then reconstructed by
summing over mode orders and inverting the spatial Fourier transform. After some
manipulation of equation~\eqnref{eqn:RayPaths} this inverse transform is expressed
as

\begin{equation} \label{eqn:IFT_Green}
	G(s,s^\prime, y-y^\prime, t-t^\prime) = \frac{H(t-t^\prime)}{2\pi} \sum_{n=1}^\infty \int_{-\infty}^\infty d\kappa \;
		U_n(s) \, U_n(s^\prime) \, \frac{\sin\left[\omega_n(\kappa) \,(t-t^\prime)\right]}{\omega_n(\kappa)} \; e^{i\kappa (y-y^\prime)} \; .
\end{equation}

\noindent where $H$ is the Heaviside step function. The inverse transform has an
analytic solution \citep{Weast:1989} involving zero-order Bessel functions of the
first kind, $J_0$,

\begin{equation} \label{eqn:GreenF}
	G(s,s^\prime; y-y^\prime, t-t^\prime) = \frac{H(\tau)}{2\va} \sum_{n=1}^\infty
		U_n(s) U_n(s^\prime) \, J_0\left(\lambda_n \va T\right)  \; .
\end{equation}

\noindent We have written the solution compactly by defining the following
delayed times:

\begin{eqnarray}
	\tau &\equiv& (t-t^\prime) - \frac{\left|y-y^\prime\right|}{\va} \; ,
\\ \nonumber \\
	T &\equiv& \sqrt{(t-t^\prime)^2 - \frac{(y-y^\prime)^2}{\va^2}} \; .
\end{eqnarray}

\noindent As expected, the Green's function is nonzero only after the excitation
occurs and after the first wave fronts arrive at the observation point ($s$,$y$),
i.e., when $\tau > 0$. Further, since both the source and the observation point are
within a waveguide, there are many paths that waves can take from the point source
at $y^\prime$ to the observation point at $y$, each reflecting a different number
of times from the walls of waveguide (in this case the footpoints). The superposition
of waves traveling along all these paths generates the oscillation pattern seen at
any given point. This superposition is a combination of waves with different wavenumbers
$\kappa$ and hence directions of initial launch from the source. This summation is
represented by the integral in the inverse Fourier transform in equation~\eqnref{eqn:IFT_Green}.

The response of the waveguide to the impulsive source appearing in
equation~\eqnref{eqn:TwoSource}, $S_{\rm imp}(s)\delta(t-t^\prime)\delta(y-y^\prime)$
is of course the integral of the product of the Green's function and the source
$S_{\rm imp}(s^\prime)$ over the point source's location $s^\prime$,

\begin{eqnarray}
	v_{\rm imp}(s,y,t) &=& \frac{H(\tau)}{2\va} \sum_{n=1}^\infty {\cal A}_n \, U_n(s) \, J_0\left(\lambda_n \va T\right)  \;
\\ \nonumber \\
	{\cal A}_n &\equiv& \int_0^L S_{\rm imp}(s^\prime) U_n(s^\prime)\, ds^\prime \;  .
\end{eqnarray}

\begin{figure*}%
        \epsscale{1.0}%
        \plotone{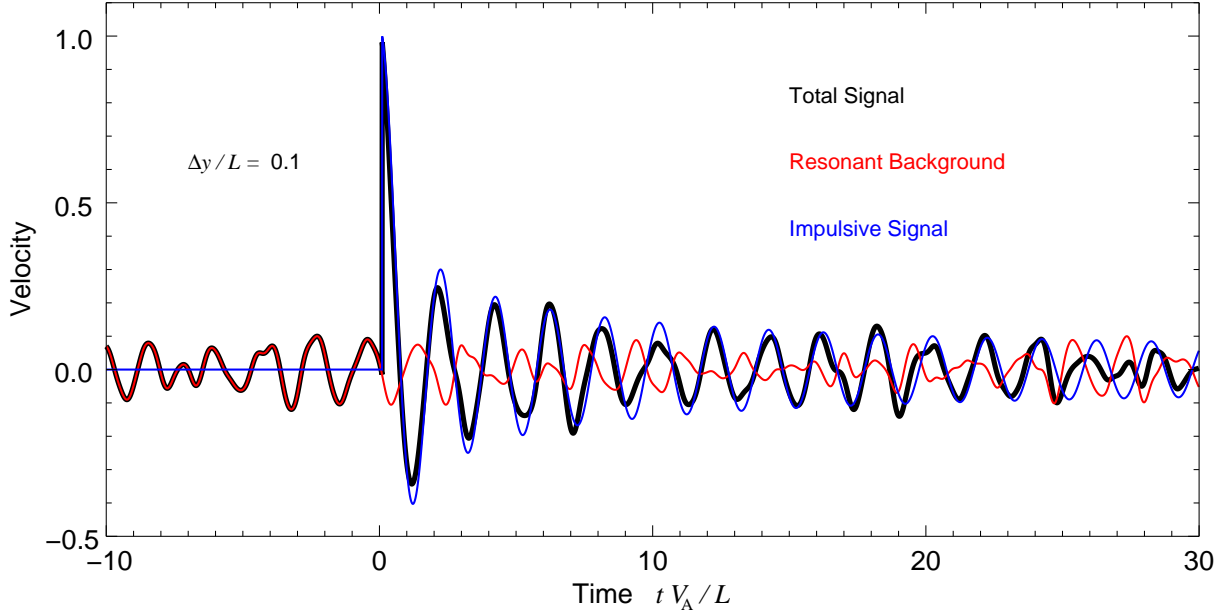}%
        \caption{\small The wave signal as a function of time arising from both
wave sources. The red curve is the background signal generated by a distributed
stochastic source. The strength of the stochastic source is gaussian in wavenumber
$\kappa$ (see Figure~\ref{fig:GaussSpec}$a$). The dominant frequency is the lowest
frequency available, corresponding to those waves with $\kappa = 0$ which propagate
parallel to the magnetic field. The blue curve is the signal arising
from a single impulsive event that occurred rather close to the observation point
along the arcade, $\Delta y = 0.1~L$. The initial pulse corresponds to waves that
have propagated straight down the waveguide, while the latter oscillations are an
interference pattern arising from waves that have taken a variety of paths down the
waveguide. The black curve shows the total wave signal. We have chosen the relative
size of the impulsive and background sources such that the background has an amplitude
of 10\% of the initial pulse height of the impulsive signal.
\label{fig:WaveSigG0.1}}%

\end{figure*}

\begin{figure*}%
        \epsscale{1.0}%
        \plotone{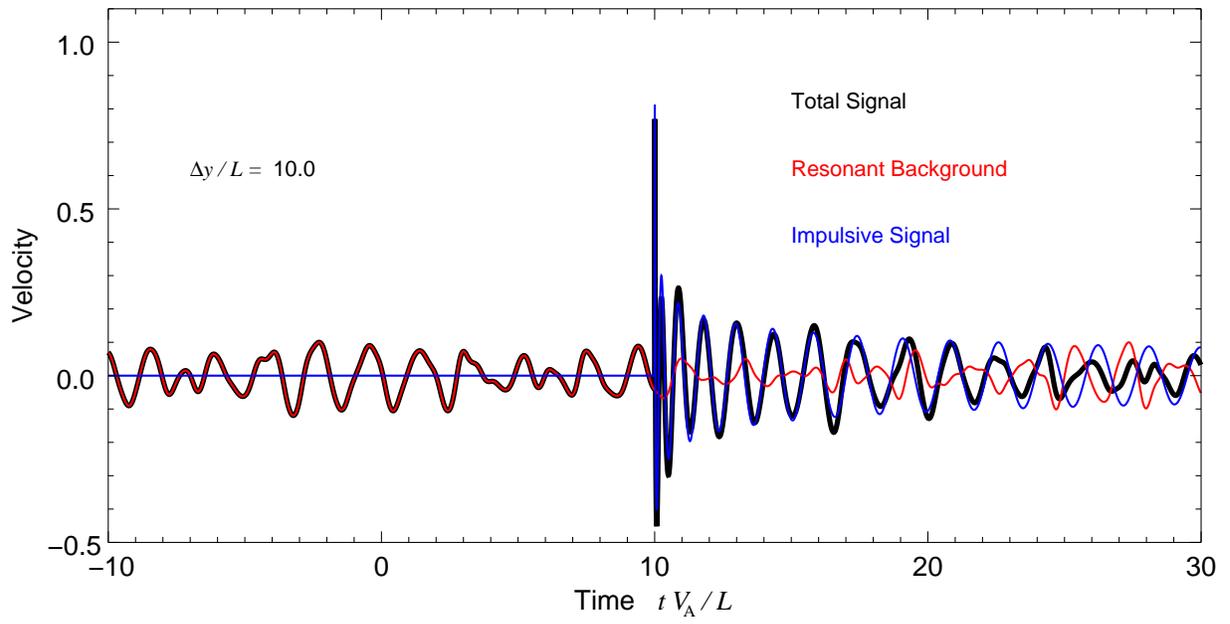}%
        \caption{\small As in Figure~\ref{fig:WaveSigG0.1}, except the observation
point is much farther from the impulsive source, $\Delta y = 10~L$. In addition to
the existence of the expected delay required for the waves to arrive at the observation 
point, the fringes in the interference pattern become compressed in time near the
time of first arrival.
\label{fig:WaveSigG10}}%

\end{figure*}

\begin{figure*}%
        \epsscale{1.0}%
        \plotone{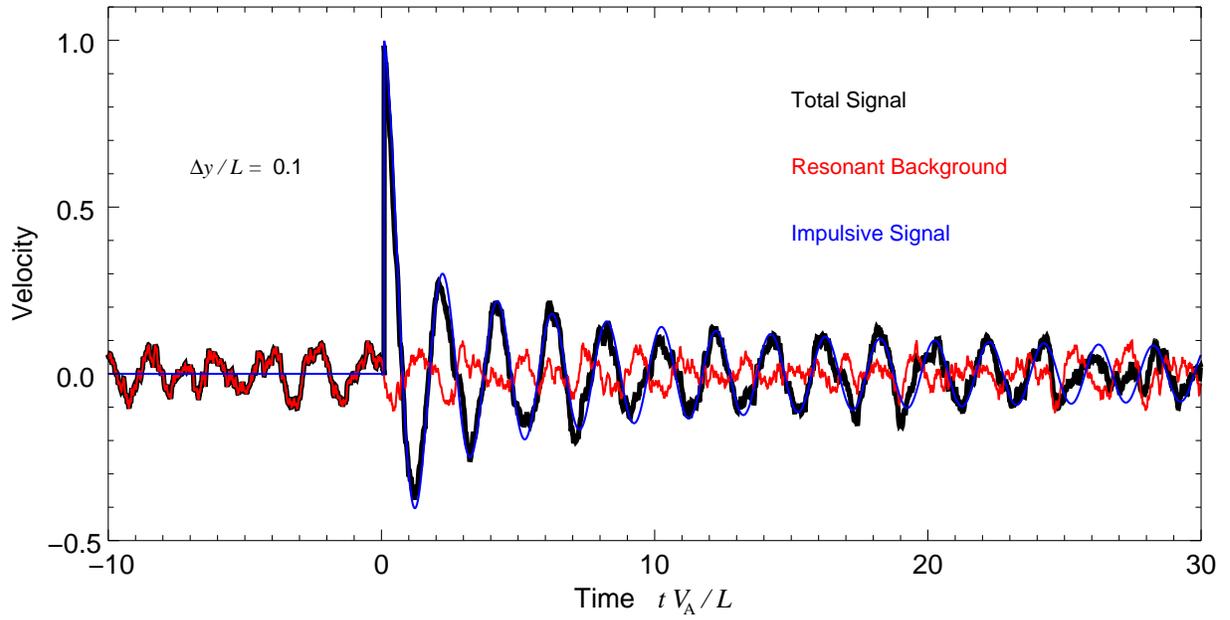}%
        \caption{\small As in Figure~\ref{fig:WaveSigG0.1}, except the background
source is white with equal power in all wavenumbers. The resulting background signal
is still dominated by the lowest waveguide frequency, but the relative importance
of high frequency waves is clearly visible.
\label{fig:WaveSigW}}%

\end{figure*}

In Figures~\ref{fig:WaveSigG0.1}--\ref{fig:WaveSigW} we show this signal superimposed
on the background oscillations. In all cases, the impulsive source occurs at time $t=0$
and the waves are observed at the apex of the loop $s = L/2$. Further, for simplicity
we assume  that only the gravest mode $n=1$ is excited to significant amplitude (i.e.,
$|{\cal A}_n| \ll |{\cal A}_1|$ for $n \neq 1$). Figures~\ref{fig:WaveSigG0.1}
and \ref{fig:WaveSigW} correspond to an impulsive event that occurs only a short
distance away from the observation point, $\Delta y = y-y^\prime = 0.1\; L$, while
for Figure~\ref{fig:WaveSigG10} the distance between the source and observation
point is a one-hundred times larger, $\Delta y = 10\; L$. Not only is there
increased delay between the event and reception of the first signal for the
more distant source, but the fringe pattern is compressed near the time of
first arrival. Thus, more distance sources generate signals with a wider range
of apparent frequencies.


\section{Discussion}
\label{sec:Discussion}

We have proposed an alternate model for coronal loop oscillations. Instead of the
standard picture that the visible loop is a self-contained 1D oscillator, we propose
that the observed waves are MHD fast waves that live on the entire arcade and are
inherently 2D in nature. The waves are trapped longitudinally between the loci of
field line footpoints in the photosphere, but freely propagate along the axis of
the arcade perpendendicular to the field lines. Therefore, the arcade forms a
2D waveguide with modes that have discretized wavenumbers in the longitudinal
direction $\lambda_n$ and continous wavenumbers $\kappa$ in the axial direction.

We demonstrate that both the ``decaying" flare-induced oscillations and the low-amplitude
``decayless" oscillations that have been observed \citep{Nistico:2013, Anfinogentov:2013}
can be explained by such 2D waves if there are two distinct wave sources: a continuous,
distributed, stochastic source and a large-amplitude impulsive source, localized both
spatially and temporally. For this model, the inclusion of a physical damping mechanism
(such as phase mixing or resonant absorption) is not necessary to reproduce the general
behavior of either observed wave component. We discuss the properties of the wavefield
excited by both of these sources in the following subsections.


\subsection{Decayless Oscillations}
\label{subsec:Decayless}

The decayless oscillations seen by \cite{Nistico:2013} and \cite{Anfinogentov:2013}
appear to be reproduced with fidelity by considering the effect of a stochastic
source that operates throughout the waveguide and continues for long durations.
Such a source produces a profusion of waveguide modes with uncorrelated phases.
For each longitudinal order $n$, these modes form a continuous spectrum in the
transverse wavenumber $\kappa$ and therefore frequency $\omega$. Each spectrum
has a low frequency cut-off below which no modes exist (see Figure~\ref{fig:PowSpec}).
This cut-off corresponds to modes that propagate parallel to the field lines and
hence do not travel up and down the waveguide. As such, these modes are those
that are most analogous to those that would be obtained in a 1D model where
one assumes that thin bundles of field lines are individually resonant.

For a source with wavenumber dependence that is sufficiently flat near $\kappa = 0$,
i.e., near the cut-off, we expect that most of the relevant modes saturate such that
they have equal energy. Therefore, since the mode energy is proportional to the 
square of the velocity amplitude and to the square of the frequency, the mode amplitude
should be inversely proportional to the mode frequency. This means that the dominant
frequency in the spectrum should be the low-frequency cut-off. Thus, in observations
we should expect to primarily see the fundamental ($n=1$) waveguide mode that
propagates nearly parallel to the field lines ($\kappa = 0$). This is, of course,
exactly what is observed for the large-amplitude flare-induced waves \citep{Aschwanden:1999,
Nakariakov:1999}, but has yet to be verified for the low-amplitude decayless
oscillations.

We further point out that while the spectrum of oscillations is dominated by
the mode with the lowest frequency, the wavefield also contains higher frequency
components. The relative importance of the high-frequency waves depends on the
spectral content of the source. White spectra produce noticeable high-frequency
jitter (see Figure~\ref{fig:WhiteSpec}) whereas a more narrow-band source
has a smoother response (see Figure~\ref{fig:GaussSpec}). The signal with
high-frequency jitter is quite reminiscent of the decayless oscillations
presented by \cite{Nistico:2013}. Furthermore, the beating and slow modulation
of the phase caused by interference between different nearby frequency components
is also seen in these observations.


\subsection{Flare-Induced Oscillations}
\label{subsec:FlareInduced}

A point source located within the waveguide generates a circular wavefront that
initially expands isotropically in two-dimensions across the arcade's magnetic
sheet.  This isotropic expansion stops, when the wavefront impacts the photosphere
and reflection occurs. These reflections then begin to interfere with other portions
of the wave front and after many reflections the interference pattern can become
rather complicated. Observations made some distance down the waveguide from the
point source will see an oscillatory fringe pattern produced by this interference. Thus,
the oscillation signal that is seen does not arise from a resonance occuring on
the field line where the observation is made. Instead, the oscillation is an
interference pattern of many waves as they propagate past the observation point. 
The initial pulse arises from the segment of the wave front that propagated straight
down the waveguide without reflection (i.e., waves with $\kappa \gg \lambda_n$).
At later times, the signal is the interference of segments of the initial wave front
that have taken different paths down the waveguide, all with the same path length.
As time passes the waves that arrive have undergone more and more reflections and
therefore have smaller and smaller wavenumber $\kappa$. Asymptotically, for very
long times all waves contributing to the signal have $\kappa \ll \lambda$ and thus
nearly identical frequencies of $\omega_n = \lambda_n \va$. Thus, the signal stabilizes
to the same frequency that one would obtain for a 1D cavity.

One important consequence is that the signal at the observation point decays, but
it does not do so because of physical damping. We have not included any dissipation 
mechanisms in our model. Because of this the decay does not have the exponental
fall off with time as one would expect from physical damping. Instead, as indicated
by the asymptotic form of the $J_0$ Bessel function, the signal decreases with time
like a power law $1/t^{1/2}$.  This decay rate (and the fringe pattern itself) is a
direct consequence of the shape of the waveguide and the distance from source to
observation point. The shape of the waveguide determines the possible paths and
therefore the interference. The distance between source and observation point is
important because the excited waves have differing phase speeds parallel to the
axis of the  waveguide,

\begin{equation}
	\frac{\omega}{\kappa} = \left(1 + \frac{\lambda_n^2}{\kappa^2}\right)^{1/2} \va \; .
\end{equation}

\noindent Therefore, the waves disperse as they travel down the waveguide and the
wave packet elongates and changes shape. Thus, the resulting fringe pattern depends
on how far the waves have traveled from the source. This effects manifests as the
delayed time $T=\sqrt{\Delta t^2 - \Delta y^2/\va^2}$ that appears in the argument of the
Bessel function.

Finally, we comment that not all observations of flare-driven kink waves have a
sudden onset followed by rapid day. Some appear to grow initially and only afterward
begin to decay \citep[see][]{Nistico:2013, Wang:2012}. Such cases are likely
the result of a source with a duration that is comparable to or longer than the
period of the waves that are excited. The resulting signal would be the temporal
convolution of the source with the Green's function. So, the fringe pattern that
would be observed would not only depend on the shape of the waveguide and the
distance from the source, but also the duration and temporal variation of the source
itself.


\subsection{Conclusions}
\label{subsec:Conclusions}

The interpretation of coronal-loop oscillations that we suggest here involves the
resonances of a 2D arcade instead of a 1D loop. Therefore, this new picture complicates
how mode frequencies might be extracted from an observed time series as the time series
has a richer high-frequency spectrum of waves that propagate obliquely to the field.
Fortunately, the dominant frequencies correspond to the same type of wave that one
would derive from a 1D model. Thus, these frequencies can still be used in a seismic
analysis as others have previously envisioned.

While none of our figures have included the signal from higher-frequency overtones
($n>1$), such modes will certainly be excited. Their exact amplitude depends on the
distribution of the driver along the field lines, but in all cases the amplitudes
of overtones likely decrease with mode order as high-frequency modes tend to have
lower amplitudes even for modes with equal energy. We wish to point out that if
one is attempting to measure overtone frequencies from flare-induced oscillations,
one must be careful. The response of the fundamental mode of the waveguide to a
point source is polychromatic. A wavelet analysis would suggest that the frequency
of the oscillation slowly decreases from onset until an asymptotic value is achieved.
A distant source in particular may start oscillating with a rather high frequency
compared to its eventual asymptotic value (see Figure~\ref{fig:WaveSigG10}). It is
this asymptotic value that corresponds to the 1D resonant frequencies. Of course
in many observations a loop oscillation may only be visible for several cycles and
the asymptotic regime may never be reached before the flare-induced signal falls
below the background oscillations. Due to the polychromatic nature of flare-induced
oscillations, the low-amplitude decayless oscillations may be a better frequency
diagnostic as the frequency content of the signal is largely steady with time. This
property might allow significant averaging of Fourier (or wavelet) power spectra
such that the low-frequency cut-offs that should be present for each mode order
become visible and measurable.

Finally, we emphasize that that the decay of the flare-induced signal may have nothing
to do with physical damping. In our model the decay is a wave interference effect and
the resulting fringe pattern is sensitive to the shape of the waveguide. An arcade
comprised of loops with a wide variety of lengths should generate a very different
fringe pattern and concomitant decay rate than the rectangular waveguide employed
here. Further, spatial variation of the Alfv\'en speed within the waveguide will
change the raypaths that combine to form a fringe. Thus with further analysis, the
decay rate might prove to be a useful diagnostic of the wavespeed when utilized in
tandem with the frequencies.


\acknowledgements

This work was supported by NASA, RSF (University of Sheffield) and STFC (UK). BWH
acknowledges NASA grants NNX08AJ08G, NNX08AQ28G, and NNX09AB04G.

\appendix\section{Modal Expansion of the Background Signal}
\label{app:ModalExpansion}

In this appendix we provide a derivation of the wavefield generated by a stochastic
source that is distributed both spatially and temporally. We do so in a standard
way by expressing the solution as a sum over the modes of the waveguide. We begin
by considering the contribution to the wavefield that arises from the background
source $S_{\rm bg}(s,y,t)$. The wavefield generated by this source must obey
Equation~\eqnref{eqn:WaveEqnComp} with the background source appearing on the
right hand side,

\begin{eqnarray} \label{eqn:WaveEqnBG}
	\left[\dparn[ ]{t}{2} + \omega_{n}^2(\kappa)\right] \hat{v}^{\rm (bg)}_{n}(\kappa,t) &=& \hat{S}^{\rm (bg)}_{n}(\kappa,t) \; ,
\\ \nonumber \\
	\hat{S}^{\rm (bg)}_{n}(\kappa,t) &=& \int_{-\infty}^\infty dy \int_0^L ds \; S_{\rm bg}(s,y,t) \, U_{n}(s) \, e^{-i\kappa y} \; .
\end{eqnarray}

We now take the temporal Fourier transform of these equations, adopting the notation
that $f(\omega)$ is the transform of $f(t)$,

\begin{equation}
	f(\omega) = \int_{-\infty}^{\infty} dt \, f(t) \, e^{i \omega t} \; .
\end{equation}

\noindent Note, the opposite sign convention that appears in the oscillatory
waveform used in the spatial versus temporal transform. This convention was chosen
to ensure that waves with positive wavenumber $\kappa$ correspond to waves propagating
in the positive $y$ direction. After solving for the velocity amplitude, the transform
of equation~\eqnref{eqn:WaveEqnBG} produces

\begin{eqnarray}
	\hat{v}^{\rm (bg)}_{n}(\kappa,\omega) = -\frac{\hat{S}^{\rm (bg)}_{n}(\kappa,\omega)}{\omega^2 - \omega_n^2(\kappa)} \; .
\end{eqnarray}

\noindent The solution expressed in time $t$ is now obtained by inverting the temporal
Fourier transform,

\begin{eqnarray}
	\hat{v}^{\rm (bg)}_{n}(\kappa,t) = -\frac{1}{2\pi} \int_{-\infty}^\infty d\omega \,  \frac{\hat{S}^{\rm (bg)}_{n}(\kappa,\omega)}{\omega^2 - \omega_n^2(\kappa)}\, e^{-i \omega t} \; .
\end{eqnarray}

This integral can be evaluated by contour integration. Assuming that the source is
analytic and lacks poles or continuous spectra, when the contour is deformed downwards
in the complex-frequency plane the contribution from the modes is picked up as the
residues around the poles of the integrand. There are two poles, one for positive
frequencies and the other for negative frequencies, each corresponding to waves
propagating in opposite directions up and down the waveguide,

\begin{eqnarray}
	\hat{v}^{\rm (bg)}_{n}(\kappa,t) = \frac{1}{2i\omega_n(\kappa)} \left[\hat{S}^{\rm (bg)}_{n}(\kappa,\omega_n) \, e^{-i\omega_n t} - 
		\hat{S}^{\rm (bg)}_{n}(\kappa,-\omega_n) \, e^{i\omega_n t} \right] \; .
\end{eqnarray}

\noindent For the sake of clarity we have momentarily dropped the explicit $\kappa$
dependence from the mode frequencies that appear in both the exponentials and the
source function.

We now transform back into configuration space by inverting the spatial transform
and summing over waveguide modes,

\begin{eqnarray}
	v_{\rm bg}(s,y,t) = \frac{1}{2\pi} \sum_{n=1}^\infty \int_{-\infty}^\infty d\kappa \; 
		\left[\frac{\hat{S}^{\rm (bg)}_{n}(\kappa,\omega_n)}{2i\omega_n(\kappa)} \, e^{-i\omega_n t}  -
		\frac{\hat{S}^{\rm (bg)}_{n}(\kappa,-\omega_n)}{2i\omega_n(\kappa)} \, e^{i\omega_n t} \right] \, U_n(s) \, e^{i \kappa y}\; .
\end{eqnarray}

\noindent We can put this integral in a more convenient form by making a change of variable
in the integral represented by the second term in the square brackets, $\kappa \to -\kappa^\prime$,
and then changing the name of the dummy variable back to the original, $\kappa^\prime = \kappa$.
Noting that the eigenfrequencies are symmetric, $\omega_n(-\kappa) = \omega_n(\kappa)$, we
obtain

\begin{eqnarray} \label{eqn:BG-almost}
	v_{\rm bg}(s,y,t) = \frac{1}{2\pi} \sum_{n=1}^\infty  \int_{-\infty}^\infty d\kappa \, 
		U_n(s) \, \left[\frac{\hat{S}^{\rm (bg)}_{n}(\kappa,\omega_n)}{2i\omega_n(\kappa)} \, e^{i(\kappa y - \omega_n t)} - 
		\frac{\hat{S}^{\rm (bg)}_{n}(-\kappa,-\omega_n)}{2i\omega_n(\kappa)} \, e^{-i(\kappa y -\omega_n t)} \right] \; .
\end{eqnarray}

We now use the fact that the source function (in configuration space) is
a real function. Therefore, its transform has complex-conjugate symmetry,

\begin{equation}
	\hat{S}^{\rm (bg)}_n(-\kappa,-\omega) = \left[\hat{S}^{\rm (bg)}_n(\kappa,\omega)\right]^* \; ,
\end{equation}

\noindent Using this symmetry property, we can rewrite equation~\eqnref{eqn:BG-almost}

\begin{eqnarray}
	v_{\rm bg}(s,y,t) = \frac{1}{2\pi} \sum_{n=1}^\infty \int_{-\infty}^\infty d\kappa \;
		\frac{\left|\hat{S}^{\rm (bg)}_{n}(\kappa,\omega_n)\right|}{\omega_n(\kappa)} \,
		U_n(s) \, \sin\left[\kappa y - \omega_n(\kappa) t + \theta_n(\kappa)\right] \; ,
\end{eqnarray}

\noindent where we have defined the complex phase of the source function evaluated at
the mode frequencies,

\begin{equation}
	\theta_n(\kappa) \equiv \arg\left\{\hat{S}^{\rm (bg)}_{n}(\kappa,\omega_n)\right\} \; .
\end{equation}


\section{Calculation of the Green's Function}
\label{app:GreensFunc}

The Green's function is of course the response of the system to a single point source
of unit amplitude. Therefore, consider such a source that occurs at time $t=t^\prime$ 
and at location $(s,y) = (s^\prime, y^\prime)$,

\begin{equation}
	S(s,y,t) = \delta(s-s^\prime) \, \delta(y-y^\prime) \, \delta(t-t^\prime)  \;
\end{equation}

\noindent The Fourier transform of such a source has the following modal decomposition,

\begin{equation}
	\hat{S}_n(\kappa,t) =  U_n(s^\prime) \, e^{-i\kappa y^\prime} \, \delta(t-t^\prime) \; .
\end{equation}

If we insert this expression into the right hand side of equation~\eqnref{eqn:WaveEqnComp}
we obtain an equation that describes the temporal evolution for each component in the
decomposition of the Green's function,

\begin{equation} \label{eqn:WaveEqnGreen}
	\left[\dparn[ ]{t}{2} + \omega_n^2(\kappa)\right] \hat{G}_n(s^\prime, \kappa, \Delta t) =
		U_n(s^\prime) \, e^{-i\kappa y^\prime} \, \delta(\Delta t) \; ,
\end{equation}

\noindent where we have defined $\Delta t \equiv t - t^\prime$. At the time of
the excitation event $t = t^\prime$ (or $\Delta t = 0$) the solution must satisfy
appropriate jump conditions,

\begin{eqnarray}
	\left[\hat{G}_n\right]_{\Delta t = 0} &=& 0\; ,
\\ \nonumber \\
	\left[\dpar[\hat{G}_n]{t}\right]_{\Delta t = 0} &=& U_n(s^\prime) \, e^{-i\kappa y^\prime} \; .
\end{eqnarray}

\noindent The well-known solution is a sinusoid times a Heaviside step function $H$,

\begin{equation}
	\hat{G}_n(s^\prime, \kappa,\Delta t) = U_n(s^\prime) \, e^{-i\kappa y^\prime} \, H(\Delta t) \, \frac{\sin\left[\omega_n(\kappa) \Delta t\right]}{\omega_n(\kappa)}  \; .
\end{equation}

\noindent Because of the particular functional form of $\omega_n(\kappa)$,

\begin{equation}
	\omega_n(\kappa) = \va \left(\lambda_n^2 + \kappa^2\right)^{1/2} \; ,
\end{equation}

\noindent the inverse Fourier transform of the Green's function has a standard solution
\citep{Weast:1989} which demonstrates that information travels at a finite speed, i.e.,
the Alfv\'en speed $\va$,

\begin{eqnarray}
	G_n(s^\prime, \Delta y,\Delta t) &=& \frac{1}{2\pi} \int_{-\infty}^\infty d\kappa \;
		G_n(s^\prime, \kappa,\Delta t) \, e^{i\kappa y} \; ,
\\ \nonumber \\
		\label{eqn:RayPaths}
		&=& \frac{1}{2\pi} \, U_n(s^\prime) \, H(\Delta t) \int_{-\infty}^\infty d\kappa \; 
		\frac{\sin\left[\omega_n(\kappa) \Delta t\right]}{\omega_n(\kappa)} \, e^{i\kappa \Delta y} \; ,
\\ \nonumber \\
		\label{eqn:Gn}
		 &=& \frac{1}{2\va} \, U_n(s^\prime)\, H(\Delta t) \, H(\tau) \, J_0\left(\lambda_n \va T\right) \; ,
\end{eqnarray}

\noindent where we have made the following definitions,

\begin{eqnarray}
	\tau &\equiv& \Delta t - \frac{\left|\Delta y\right|}{\va} \; ,
\\ \nonumber \\
	T &\equiv& \sqrt{\Delta t^2 - \Delta y^2/\va^2} \; ,
\\ \nonumber \\
	\Delta y &\equiv& y - y^\prime \; .
\end{eqnarray}

\noindent In this solution, the function $J_0$ is the zeroth-order Bessel function
of the first kind. Since the product of Heaviside step functions in equation~\eqnref{eqn:Gn}
is nonzero only if $\tau > 0$, the Green's function has the following solution in
configuration space,

\begin{equation} \label{eqn:GreenFunc}
	G(s,s^\prime, \Delta y, \Delta t) = \frac{H(\tau)}{2\va} \sum_{n=1}^\infty
		U_n(s) U_n(s^\prime) \, J_0\left(\lambda_n \va T\right)  \; .
\end{equation}




\end{document}